\pdfoutput=1
\documentclass[aps,twocolumn,a4paper,showpacs,superscriptaddress,
    floatfix]{revtex4-1}
\usepackage{amsmath,amssymb}
\usepackage[hypertexnames=false]{hyperref}
\usepackage{datetime}
\usepackage{enumitem}
\usepackage{graphicx}
\usepackage{braket}
\usepackage{esdiff}
\usepackage{MnSymbol}

\usepackage{siunitx}
\usepackage[dvipsnames]{xcolor}
\usepackage{xfrac} %sfrac
\usepackage{tikz}

\DeclareSIUnit\gauss{G}

\hypersetup{
    colorlinks,
    linkcolor={red!50!black},
    citecolor={blue!50!black},
    urlcolor={blue!80!black}
}

\begin{document}

\title{Quantized conductance through a spin-selective atomic point contact}
\author{Martin Lebrat}
\author{Samuel H\"{a}usler}
\author{Philipp Fabritius}
\author{Dominik Husmann}
\author{Laura Corman}
\author{Tilman Esslinger}
\affiliation{Department of Physics, ETH Zurich, 8093 Z\"{u}rich, Switzerland}

% \date{\today, \currenttime}

\begin{abstract}
We implement a microscopic spin filter for cold fermionic atoms in a quantum point contact (QPC) and create fully spin-polarized currents while retaining conductance quantization.
Key to our scheme is a near-resonant optical tweezer inducing a large effective Zeeman shift inside the QPC while its local character limits dissipation.
We observe a renormalization of this shift due to interactions of a few atoms in the QPC.
Our work represents the analog of an actual spintronic device and paves the way to studying the interplay between spin-splitting and interactions far from equilibrium.
\end{abstract}

\maketitle

\begin{figure}
    \includegraphics{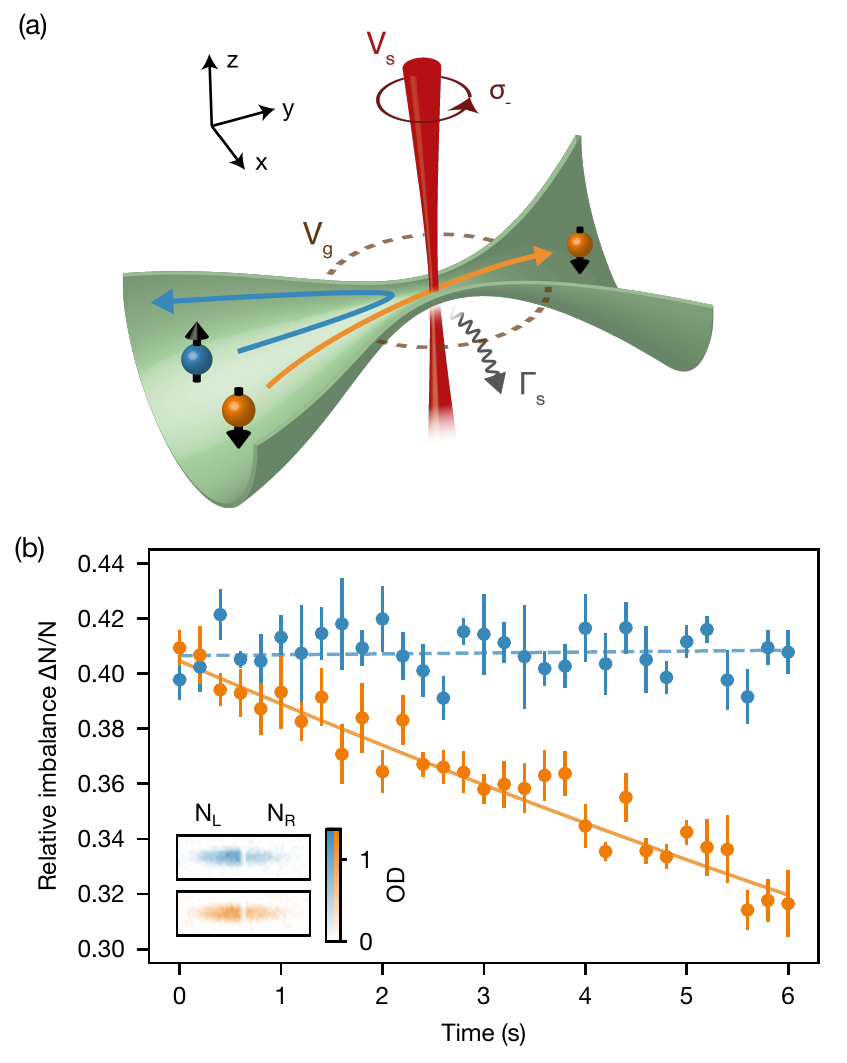}
    \caption{\textbf{Creating spin-polarized currents through an atomic QPC.} (\textbf{a}) A near-resonant optical tweezer (red) with waist $w_s = \SI{2.0(1)}{\micro\meter}$ introduces an effective Zeeman shift $V_s$ inside an optically-defined QPC (green).
    It allows $^6$Li atoms in the lowest hyperfine state ($\ket{\downarrow}$, orange) to flow between two reservoirs while blocking atoms in the third-lowest state ($\ket{\uparrow}$, blue) thereby acting as a spin filter with losses determined by the photon scattering rate $\Gamma_s$.
    A far-detuned gate beam (dashed circle) locally increases the chemical potential $\mu_\text{res}$ imposed by the reservoirs by $V_g$.
    (\textbf{b}) Time evolution of the relative atom number difference between the left and right reservoirs $\Delta N / N = (N_L - N_R)/(N_L + N_R)$, obtained for a mean chemical potential $V_g + \mu_\text{res} = k_B \cdot \SI{0.61(2)}{\micro\kelvin}$, a near-resonant beam power $P_s = \SI{20(6)}{\pico\watt}$ and at a scattering length $a = 0(7)\,a_0$.
    It is constant for $\ket{\uparrow}$ and has a decay time of $\SI{25(1)}{\second}$ for $\ket{\downarrow}$.
    The associated currents $I_\uparrow = -19 \pm 85$ atoms/s and $I_\downarrow = 833 \pm 98$ atoms/s indicate fully polarized transport within fit error.
    Error bars show the standard error of the mean of 5 measurements here and in Fig.~\ref{fig:evolution}.
    Inset: Optical density (OD) after $\SI{6}{\second}$ of left and right reservoirs over $970 \times \SI{320}{\micro\meter}$, averaged over 5 absorption images.
    }
    \label{fig:fig1}
\end{figure}

Coupling the spin of a particle to its motion unveils its quantum nature, as was demonstrated in the Stern-Gerlach experiment \cite{gerlach_experimentelle_1922}.
In condensed-matter systems, this can be achieved via Zeeman effect or spin-orbit coupling and gives rise to a variety of transport phenomena.
For instance, magnetic impurities coupled to a metallic or superconducting bath strongly influence resistivity \cite{kondo_resistance_1964,shiba_classical_1968}, and spin-orbit coupling can induce spin-polarized modes and Majorana fermions at the edge of topological materials \cite{kane_$z_2$_2005,beenakker_search_2013}.

Cold atoms provide an alternative platform to investigate spin transport with long coherence times and a fine tuning of interactions by encoding spin into different hyperfine states.
There, spin degeneracy can be lifted with actual magnetic fields or differential Stark shifts \cite{jessen_optical_1996,mandel_coherent_2003,steffen_digital_2012}, and spin-orbit coupling can be realized using Raman schemes \cite{lin_spinorbit-coupled_2011,wang_spin-orbit_2012,cheuk_spin-injection_2012}.
Experimental realizations have so far addressed spin-dependent effects on a global scale.
However, a local manipulation of the spin would allow the study of interfaces, as encountered in magnetic heterojunctions, and in general represents a central ingredient for spintronics \cite{zutic_spintronics:_2004}, quantum computation \cite{loss_quantum_1998} and quantum simulation \cite{jane_simulation_2003}.

In this work, we use a quantum gas experiment \cite{krinner_two-terminal_2017} to probe spin transport through a microscopic one-dimensional channel connected to two macroscopic reservoirs.
By focusing a near-resonant optical tweezer inside the channel, we realize an effective Zeeman splitting which is large compared to all other transport energy scales and allows us to individually control spin currents.
As the state of the reservoirs is not affected, transport measurements are carried out around a well-defined equilibrium and we observe spin-polarized quantized conductance.
Owing to the local character of the tweezer, we reduce losses caused by photon scattering and increase our experimental timescales to several seconds.
This allows us to detect minute mean-field effects caused by merely two atoms on average in the tweezer region.
Our results are captured by an extended Landauer-B\"uttiker model whose validity is studied in a companion paper \cite{companion}, where regimes with stronger dissipation are explored as well.

We prepare a degenerate cloud of lithium-6 atoms in a balanced mixture of the first- and third-lowest hyperfine states, labeled as pseudo-spins $\ket{\downarrow}$ and $\ket{\uparrow}$, with about $N = 1.1(1) \cdot 10^5$ atoms per state at a typical temperature of $T = \SI{66\pm12}{\nano\kelvin}$.
A magnetic field is tuned close to $B = \SI{568}{\gauss}$, where collisional $s$-wave interactions between $\ket{\downarrow}$ and $\ket{\uparrow}$ vanish.
We optically imprint a quantum point contact (QPC) with transverse confinement frequencies of $\nu_x = \SI{14.0(6)}{\kilo\hertz}$ and $\nu_z = \SI{9.03(5)}{\kilo\hertz}$ by intersecting two far-detuned repulsive laser beams.
The QPC separates the cloud into two reservoirs that act as a source and drain of atoms with typical mean chemical potential $\mu_\text{res} = (\mu_L + \mu_R)/2 = k_B \cdot \SI{0.23}{\micro\kelvin}$ fixed by the total atom number.
For a non-interacting Fermi gas, the conductance per spin and transverse QPC mode with unit transmission is equal to the conductance quantum $1/h$ \cite{krinner_observation_2015}.
The number of available transport modes is set via a far-detuned attractive gate beam with Gaussian waist $w_g = \SI{31.8(3)}{\micro\meter}$.
Its peak potential $V_g$ is tunable and augments the chemical potential $\mu_\text{res}$ locally to a value $V_g + \mu_\text{res}$ as depicted in Fig.~\ref{fig:fig1}(a).

The control over each individual spin current is achieved using an additional $\sigma^-$-polarized beam centered on the QPC.
Its Gaussian intensity profile is holographically defined by a Digital Micromirror Device and has a waist of $w_s = \SI{2.0(1)}{\micro\meter}$, which is smaller than the QPC length of $\SI{5.9(1)}{\micro\meter}$ and the typical Fermi wavelength of $\lambda_F = \sqrt{h/m \nu_z} = \SI{2.7}{\micro\meter}$, where $m$ is the mass of a $^6$Li atom.
Its optical frequency $\nu_s$ is tuned between the transition frequencies of $\ket{\uparrow}$ and $\ket{\downarrow}$ to the excited manifold $^2 P_{3/2}$, inducing a repulsive dipole potential for $\ket{\uparrow}$ and an attractive potential for $\ket{\downarrow}$.
For opposite detunings relative to the two transitions, i.e. $\delta_\uparrow = - \delta_\downarrow = \SI{81.3}{\mega\hertz}$, both dipole potentials have equal magnitudes $\pm V_s$.
The magnitudes are linear in the light intensity $I_s$ for $I_s \ll I_\text{sat}$, where $I_\text{sat} = \SI{25.4}{W/m^2}$ is the saturation intensity of the transition.
The induced lightshifts can be viewed as an optical analogue to the Zeeman shift $V_s = - \mu \, B_z$ of a spin-$1/2$ particle with magnetic moment $\mu$ in a fictitious magnetic field $B_z$.
As opposed to magnetically-induced shifts, here atoms scatter photons at a rate $\Gamma_s$; this imparts kinetic energy and leads to losses.
The ratio $\Gamma_s/V_s$ is independent of $I_s$ and equal to $\SI{9.4e3}{\second^ {-1}/(k_B\,\micro\kelvin)}$ for the detuning mentioned above \cite{companion}.

Starting with equal chemical potential biases across the QPC for both spins, we apply the spin-dependent optical potential to create a spin-polarized current.
Typically, we prepare for each spin state atom number differences of $\Delta N(0) = 45(3) \cdot 10^3$ between the two reservoirs and measure their time evolution towards equilibrium.
During that time, the $s$-wave scattering length is set to $a = 0(7) \, a_0$, where $a_0 = \SI{52.9}{\pico\meter}$ is the Bohr radius, and the optical power of the near-resonant tweezer is $P_s = \SI{20(6)}{\pico\watt}$, corresponding to a peak intensity $I_s = 2 P_s/\pi w_s^2 = \SI{3(1)}{\watt/\meter^2} = 0.13(4)\,I_\text{sat}$.
Over \SI{6}{\second}, the relative atom number difference $\Delta N(t)/N(t)$ remains constant for spin $\ket{\uparrow}$ and is reduced by about one quarter for spin $\ket{\downarrow}$ with a fitted decay time of $\tau_\downarrow = \SI{25(1)}{\second}$ [Fig.~\ref{fig:fig1}(b)].
Assuming linear response, we infer currents for each spin of $I_\uparrow = -19 \pm 85$ atoms/s and $I_\downarrow = 833 \pm 98$ atoms/s.

Our scheme thus represents the cold-atom equivalent of a spin filter, a fundamental building block for spintronics previously realized in spin-polarized tunnel junctions \cite{meservey_spin-polarized_1994}, quantum point contacts under strong magnetic fields \cite{rossler_transport_2011} or double-stranded DNA illuminated by polarized light \cite{gohler_spin_2011}.
The current polarization obtained here is comparable within fitting error to the best values obtained with magnetic heterostructures \cite{marrows_spin-polarised_2005}.

\begin{figure*}
    \includegraphics{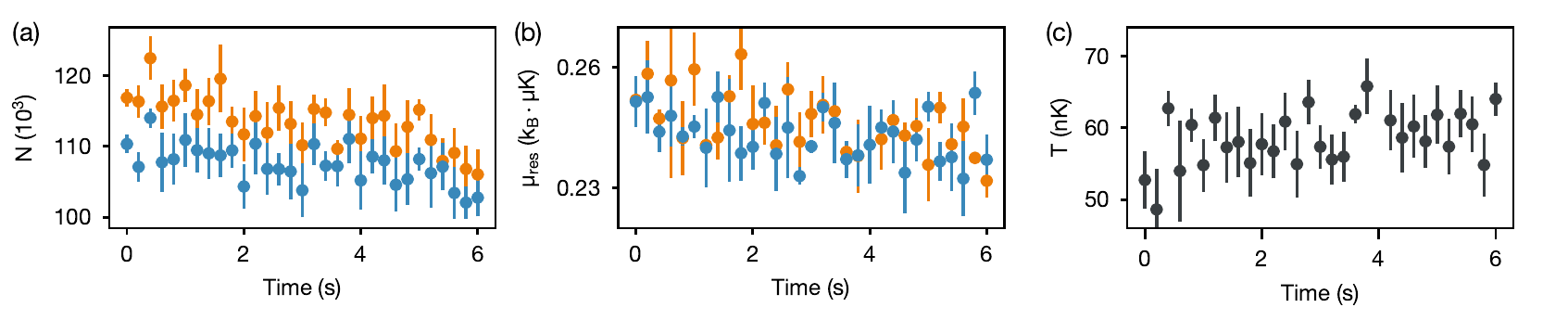}
    \caption{\textbf{Evolution of the reservoir thermodynamical quantities.}
    (\textbf{a}) Losses in atom number $N = N_L + N_R$ lead to a small variation of (\textbf{b}) mean reservoir chemical potential $\mu_\text{res} = (\mu_L+\mu_R)/2$.
    (\textbf{c}) Mean temperature $T = (T_L+T_R)/2$, constant over experimental timescales.
    }
    \label{fig:evolution}
\end{figure*}

Despite a maximal photon scattering rate $\Gamma_s = \SI{3(1)e3}{\second^{-1}}$ at the center of the tweezer, the decrease in total atom number $N$ is limited [Fig.~\ref{fig:evolution}(a)].
Losses lead to an overall decrease of the mean chemical potential $\mu_\text{res} = (\mu_L+\mu_R)/2$ in the reservoirs by $k_B \cdot \SI{30}{\nano\kelvin}$ [Fig.~\ref{fig:evolution}(b)], much smaller than the other typical energy scales.
Meanwhile, no significant increase of temperature $T$ is observed [Fig.~\ref{fig:evolution}(c)].
Since the atom-atom mean free path is larger than the system's size, the additional recoil energy $E_R = (h/\lambda)^2/2m = k_B \cdot \SI{3.54}{\micro\kelvin}$ imparted to atoms scattered by near-resonant photons with wavelength $\lambda = \SI{671}{\nano\meter}$ is not deposited in the reservoirs through thermalization.
As currents and losses are small relative to the global atom number, we however expect the reservoirs to remain effectively described by thermal states \cite{companion}. 

\begin{figure}
    \includegraphics{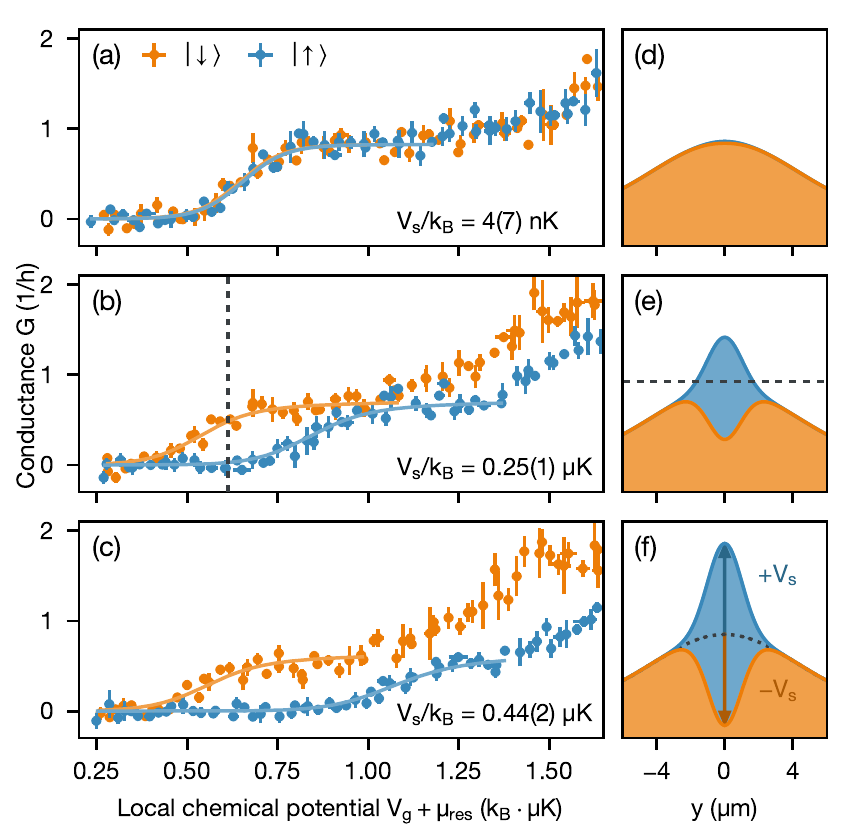}
    \caption{\textbf{Lifting the spin degeneracy of the QPC ground state.}
    (\textbf{a}) Conductance $G$ of each spin state at scattering length $a = 91(7)\,a_0$ versus local chemical potential $V_g+\mu_{\textrm{res}}$ without near-resonant beam; (\textbf{b}), with peak intensity $I_s = 0.13(4)\,I_\text{sat}$ (as in Figs.~\ref{fig:fig1} and \ref{fig:evolution}), where $I_\text{sat}$ is the $D_2$-line saturation intensity; (\textbf{c}) with $I_s = 0.17(5)\,I_\text{sat}$.
    Here and in the following, error bars correspond to the standard error of the mean of 3 measurements.
    Fits by a Landauer model are shown as solid curves and indicate an increase of the spin-dependent potential up to $V_s = k_B \cdot \SI{0.44(2)}{\micro\kelvin}$.
    (\textbf{d}), (\textbf{e}) and (\textbf{f}): Quasi-1D potentials along the transport direction $y$.
    The chemical potential of Figs.~\ref{fig:fig1} and \ref{fig:evolution} is indicated as dashed lines in (b) and (e).
    }
    \label{fig:spin_gate}
\end{figure}

Since the macroscopic state of reservoirs is only weakly affected by losses, transport occurs around a well-defined equilibrium and conductance is related to the single-particle transmission through the QPC according to the Landauer-B\"uttiker formula \cite{datta_electronic_1995}.
Probing the transmission in a spin- and energy-dependent way allows in turn to estimate the Zeeman shift induced by the near-resonant beam.
This can be done by tuning the local chemical potential $V_g+\mu_\text{res}$ relative to the potentials experienced by both spins.
We perform several conductance measurements at a weak scattering length of $a = 91(7)\,a_0$ for different values of the gate potential $V_g$.
The conductance is inferred from the decay of the relative atom number difference $\Delta N/N$ after a fixed transport time of $\SI{4}{\second}$ \cite{supplementary}.
If both spins are degenerate, we observe overlapping conductance plateaus which are characteristic of single-mode transport [Fig.~\ref{fig:spin_gate}(a)].
Their value of $G = 0.84(1)/h$, slightly below the conductance quantum $1/h$, results from a reduction of the chemical potential bias between the reservoirs due to residual temperature differences, caused by our initial preparation and omitted in the estimation of the conductance \cite{supplementary}.

Upon increasing the near-resonant beam intensity $I_s$, the conductance plateau for the repelled spin $\ket{\uparrow}$ is shifted towards larger chemical potentials [Fig.~\ref{fig:spin_gate}(b) and (c)].
This shift corresponds to the classical barrier $+V_s$ added to the QPC zero-point energy [Fig.~\ref{fig:spin_gate}(d)] by the spin-dependent potential [Fig.~\ref{fig:spin_gate}(e) and (f), blue].
An opposite shift is observed for the attracted spin $\ket{\downarrow}$, indicating a weak decrease of the potential barrier due to the near-resonant beam being smaller than the QPC length [Fig.~\ref{fig:spin_gate}(e) and (f), orange].
Current polarization is maximal for chemical potentials located between the potential barriers of both spins as in Fig.~\ref{fig:fig1}; the value of $V_g + \mu_\text{res} = k_B \cdot \SI{0.61(2)}{\micro\kelvin}$ chosen there is shown by a dashed line in Fig.~\ref{fig:spin_gate}(b) and (e).
Strikingly, plateaus persist when the intensity $I_s$ is increased to $0.17(5)\,I_\text{sat}$ while their value decreases down to $G = 0.55(2)/h$.

We expect transport observables such as conductance to be fundamentally robust against losses since they are only sensitive to scattering at energies close to the Fermi level which concerns a small fraction of all atoms subject to near-resonant light.
In a Landauer picture valid for weak interactions, these losses contribute to decreasing the conductance by the scattering probability.
This probability is equal to about 25\% at $I_s = 0.13(4)\,I_\text{sat}$ for a typical Fermi velocity $v_F = \sqrt{h\nu_z/m} = \SI{2.4}{\centi\meter/\second}$ in the single-mode regime, and is compatible with the decrease of the conductance plateau from $G = 0.84(1)/h$ in Fig.~\ref{fig:spin_gate}(a) to $G = 0.72(2)/h$ in Fig.~\ref{fig:spin_gate}(b).
In contrast, losses of atoms below the Fermi level do not generate a net current since their average velocity is zero.
In the actual setup, these losses represent the majority of the losses shown in Fig.~\ref{fig:evolution}(a) (see also \cite{companion}) and only affect transport indirectly through the weak reduction of the chemical potential [Fig.~\ref{fig:evolution}(b)].

To extract the spin-dependent potential $V_s$, we fit the conductances of both states with a Landauer model [solid curves in Fig.~\ref{fig:spin_gate}(a)-(c)].
The model describes the QPC and spin-dependent gate by two independent quasi-1D potentials shown in Fig.~\ref{fig:spin_gate}(d)-(f), and includes a position-dependent photon scattering rate $\Gamma(y)$ as an imaginary part $i \hbar \Gamma(y)/2$ \cite{companion}.
A linear regression on five different values of $I_s$ yields a conversion ratio $V_s/I_s = \SI{103(17)}{k_B \nano\kelvin/(W/\meter^2)}$ compatible with the theoretical value of $\SI{98(3)}{k_B \nano\kelvin/(W/\meter^2)}$.
We find a maximal potential of $V_s = k_B \cdot \SI{0.44(2)}{\micro\kelvin}$, about twice the typical Fermi energy $E_F = h \nu_z / 2 = k_B \cdot \SI{0.22}{\micro\kelvin}$ in the single-mode regime.

\begin{figure}
    \includegraphics{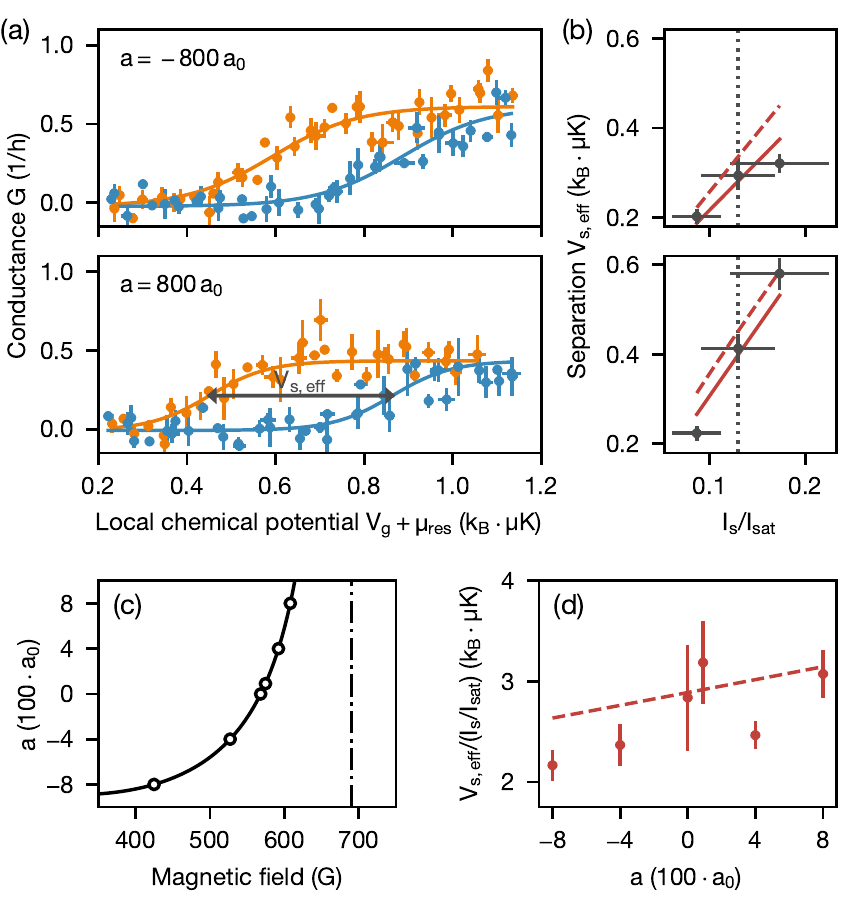}
    \caption{\textbf{Effect of 1D interactions on transport.}
    (\textbf{a}) Conductance $G$ at scattering lengths $a = -800.0(7)\,a_0$ and $a = +800(16)\,a_0$, near-resonant beam intensity $I_s = 0.13(4)\,I_\text{sat}$ and equal detuning from both states, $\bar{\delta} = 0$.
    Solid curves indicate fits with a logistic function to extract the energy separation $V_{s, \text{eff}}$.
    (\textbf{b}) Fitted separation $V_{s, \text{eff}}$ versus $I_s$ normalized by the $D_2$-line saturation intensity $I_\text{sat}$.
    The vertical dotted line indicates the intensity used in (a).
    (\textbf{c}) Scattering length $a$ versus magnetic field, tuned to the left of a broad Feshbach resonance (dash-dotted line) and calibrated within 0.1\%.
    (\textbf{d}) Ratio $V_{s, \text{eff}}/(I/I_\text{sat})$ versus scattering length $a$, obtained by linear regression displayed as solid lines in (b).
    A Hartree mean-field model including a finite temperature of \SI{66}{\nano\kelvin} is indicated as dashed lines in (b) and (d).}
    \label{fig:interactions}
\end{figure}

Using the broad Feshbach resonance of $^6$Li, we now investigate how interactions between itinerant atoms passing through the QPC compete with the Zeeman splitting created by the near-resonant tweezer.
Interactions renormalize the effective potential felt by each spin, which can be sensitively probed by measuring the energy shifts in the conductance curves.
We explore both attractive and repulsive interactions from $a = -800.0(7)\,a_0$ to $800(16)\,a_0$ in the non-superfluid regime, which are values of the scattering length where interactions are described by mean-field theory (as opposed to previous results obtained in the strongly correlated regime \cite{lebrat_band_2018}).
Figure~\ref{fig:interactions}(a) shows conductances for each spin obtained at $a = \pm 800\,a_0$ and fixed intensity $I_s = 0.13(4)\,I_\text{sat}$ as a function of the local chemical potential $V_g + \mu_\text{res}$.
We observe a change in the energy separation $V_{s, \text{eff}}$ between the conductance curves by about $k_B \cdot \SI{0.1}{\micro\kelvin}$, a shift smaller than the width of the conductance step of $4 k_B T = k_B \cdot \SI{0.26(5)}{\micro\kelvin}$.
The conductance values are reduced compared to Fig.~\ref{fig:spin_gate}(b) as well, which we attribute to slightly increased losses.
This visible separation is reduced for attractive interactions and increased for repulsive interactions.
We quantify it with a fit by a logistic function motivated by non-interacting Landauer theory for different values of $I_s$ [Fig.~\ref{fig:interactions}(b)].
The different slopes $V_{s, \text{eff}}/(I_s/I_\text{sat})$, obtained by linear regression, confirm the effect of interactions.

We repeat this measurement for intermediate scattering lengths $a$ [Fig.~\ref{fig:interactions}(c)] and extract the ratio $V_{s, \text{eff}}/(I_s/I_\text{sat})$ for each interaction strength [Fig.~\ref{fig:interactions}(d)].
The data is described by a self-consistent Hartree mean-field model \cite{supplementary} where $\ket{\downarrow}$-atoms provide an extra potential proportional to the interaction parameter $U = 2 h \sqrt{\nu_x \nu_z} a$ that ease or hinder the passage of $\ket{\uparrow}$-atoms depending on its sign.
We find a good agreement even though the model does not include dissipation and density fluctuations which may be non-negligible in the 1D region.
Crucially, our conductance signal is obtained by probing the QPC with typically a few thousand atoms over $\SI{4}{\second}$.
The mean-field approximation formally relies on replacing operators for atomic densities by their thermodynamical averages, which are about 1 atom per micron and per spin here \cite{supplementary}.
Despite the absence of thermodynamical equilibrium in the microscopic QPC, such a thermodynamical average is experimentally mimicked by the effective time average associated with our measurement.
Dissipation is likely to play a substantial role for larger interactions \cite{muller_engineered_2012,daley_quantum_2014}, where both fluctuations and coherence are expected to be stronger within the 1D region \cite{froml_fluctuation-induced_2019}.

Our work demonstrates how transport measurements are sensitive to minute interaction effects occurring on the scale of the Fermi wavelength from the integration of a weak transport signal.
Our capability to spin-engineer potentials can be readily extended to more complex structures \cite{lebrat_band_2018} and opens avenues for exploring the transport dynamics of strongly-correlated systems, where novel nonequilibrium spin and heat transport \cite{bauer_spin_2012,bergeret_colloquium:_2018} and exotic phases of matter \cite{beenakker_search_2013} could be observed.

\section*{Acknowledgments}
We thank L.~Dogra for early theoretical and technical contributions; T.~Giamarchi, L.~Glazman, H.~Moritz, H.~Ott and A.-M.~Visuri for helpful discussions; and J.-P.~Brantut, R.~Citro, M.~Landini, J.~Mohan, and K.~Viebahn for their critical reading of the manuscript.
We acknowledge the Swiss National Science Foundation (Project n$^{\circ}$ 182650 and NCCR-QSIT) and ERC advanced grant TransQ (Project n$^{\circ}$ 742579) for funding.
L.C. is supported by ETH Zurich Postdoctoral Fellowship, Marie Curie Actions for People COFUND program and EU Horizon 2020 Marie Curie TopSpiD (Project n$^{\circ}$ 746150).

% \input{refs_main.bbl}

%%% Supplementary materials %%%

\clearpage

% \setcounter{equation}{0}
% \setcounter{figure}{0}
% \makeatletter
% \renewcommand{\theequation}{S\arabic{equation}}
% \renewcommand{\thefigure}{S\arabic{figure}}
% \renewcommand{\bibnumfmt}[1]{[S#1]}
% \renewcommand{\citenumfont}[1]{S#1}

\section*{Materials and methods}

\subsection*{Experimental details}

\subsubsection*{Experimental cycle}

In brief, a balanced mixture of the lowest and third-lowest hyperfine state is loaded into a hybrid trap with transverse frequencies $\nu_{\text{trap}, x} = \SI{122(3)}{\hertz}$ and $\nu_{\text{trap}, z} = \SI{110(3)}{\hertz}$ along $x$ and $z$ provided by a $\SI{1064}{\nano\meter}$ optical dipole trap, and longitudinal frequency $\nu_{\text{trap}, y}$ along $y$ produced by a magnetic trap which ranges from 22 to $\SI{27}{\hertz}$ depending on the background magnetic field used during transport.
Typical temperatures of $\SI{66(12)}{\nano\kelvin}$ which correspond to $T/T_F = 0.24(2)$ are prepared by evaporative cooling at the $s$-wave scattering length $a = -776\,a_0$.
Subsequently the magnetic field is tuned to a value of $\SI{568}{\gauss}$ in Fig.~\ref{fig:fig1} and \ref{fig:evolution}, at $\SI{574}{\gauss}$ in Figs.~\ref{fig:spin_gate} and between $\SI{425}{\gauss}$ and $\SI{608}{\gauss}$ in Fig.~\ref{fig:interactions}.
The magnetic field is calibrated separately by measuring losses due to three-body collisions close to narrow $s$- and $p$-wave resonances whose positions are determined in \cite{schunck_feshbach_2005}.

Transport is induced by a chemical potential bias created by displacing the center of the underlying magnetic trap with a magnetic field gradient, splitting the cloud into two asymmetric reservoirs with an elliptical repulsive beam, and finally re-centering the magnetic trap.
Two additional TEM$_{01}$-shaped beams at $\SI{532}{\nano\meter}$ confine atoms at the cloud center along the $x$ and $z$ directions into a quantum point contact (QPC), with respective Gaussian waists $w_x = \SI{5.9(1)}{\micro\meter}$ and $w_z = \SI{30.2}{\micro\meter}$ in transport
direction y.
Transport is allowed through the QPC by switching off the elliptical beam for $\SI{4}{\second}$ in Figs.~\ref{fig:spin_gate} and \ref{fig:interactions}.
Atom numbers, chemical potential and temperatures for each reservoir are finally computed from the density distribution obtained from absorption pictures along the $x$-direction after a time of flight of $\SI{1}{\milli\second}$.

\subsubsection*{Shaping and alignment of the near-resonant beam}

A Digital Micromirror Device (DMD DLP7000 $0.7$'' XGA from Texas Instruments; Driver Vialux V7000) is used to precisely shape a Gaussian beam of wavelength $\lambda = \SI{671}{\nano\meter}$ onto the QPC.
The DMD is a rectangular array of $1024 \times 768$ micron-scale mirrors which can be individually controlled, thereby forming a reflective diffraction grating.
The grating imprints an amplitude and phase hologram on the near-resonant beam, whose Fourier transform is optically conjugated to the $(x, y)$ plane of the QPC with a set of relay lenses and
a high-resolution microscope.
The shape and extent of the beam are imaged using a second microscope placed symmetrically after the QPC.
Aberrations introduced by the optical setup are corrected with the DMD via a technique similar to a Hartmann-Shack analysis.

To center the near-resonant beam onto the QPC center in the $(x, y)$ plane, we increase the dissipation rate for spin $\ket{\downarrow}$ by detuning it closer to its resonance, and map out losses
as a function of the beam position, realizing a dissipative analog to a scanning gate microscope \cite{hausler_scanning_2017}.
We also ensure that parasitic diffraction orders that could reach the atoms are adequately blocked.

\subsubsection*{Intensity and spin-dependent potential calibration}

We measure the Gaussian waist $w_s = \SI{2.0(1)}{\micro\meter}$ of the near-resonant tweezer in the transport direction with the second microscope used to quantify optical aberrations.
To calibrate its intensity we precisely characterize several neutral density filters that attenuate the beam power to the picowatt range.
These filters are calibrated by imaging the beam with a camera at different powers and exposures on the one hand; and by directly measuring the decrease in beam power due to the filters at higher absolute powers on the other hand.
We deduce an attenuation by $3.4(1)$ orders of magnitude whose large uncertainty is the main source of systematic error in estimating the beam intensity.

\begin{figure}
    \includegraphics{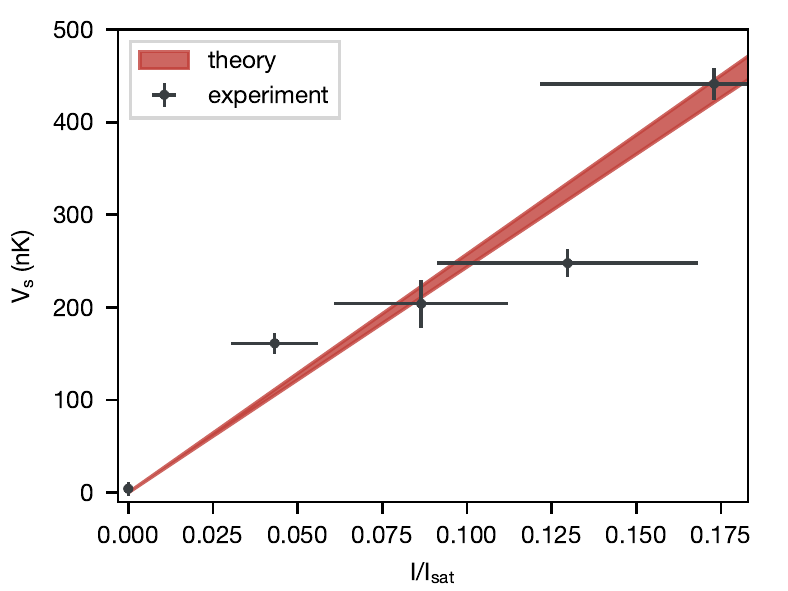}
    \caption{
    \textbf{Calibration of the spin-dependent potential.}
    Spin-dependent potential $V_s$ versus intensity $I_s$ normalized by saturation intensity $I_\text{sat}$, fitted from the conductance curves of Fig.~\ref{fig:spin_gate} (black circles) and calculated ab initio (red area).
    The horizontal and vertical error bars of the experimental points correspond to the calibration uncertainty on $I_s$ and the fit error on $V_s$ respectively.
    The error on the theory line includes calibration uncertainties on the detuning and polarization of the near-resonant beam.}
    \label{fig:spin_potential}
\end{figure}

The spin potential $V_s$ measured in Fig.~\ref{fig:spin_gate} increases with beam intensity $I_s$, as shown in Fig.~\ref{fig:spin_potential}.
As $I_s$ is small compared to the saturation intensity $I_\text{sat} = \SI{25.4}{\watt/\meter^2}$, $V_s$ is expected to be linear in $I_s$.
An orthogonal-distance regression \cite{boggs_orthogonal_1990} yields a slope of $\SI{103(17)}{\nano\kelvin/(\watt/\meter^2)}$.
It is compatible within one standard error with a theoretical value of $\SI{98(3)}{\nano\kelvin/(\watt/\meter^2)}$ independently calculated via \cite[Eq. (5)]{companion} and our knowledge of the polarization, detuning, and magnetic field.

\subsubsection*{Effective potential landscape}

Under the adiabatic approximation applied to the Landauer-B\"uttiker formalism \cite{datta_electronic_1995}, transport in the lowest transverse mode of the QPC potential is well described by the propagation of independent particles through an effective 1D potential which is the sum of the space-dependent zero-point energy due to the $x$- and $z$-confinement of the QPC,
\begin{equation}
V_0(y) =\frac 1 2  h\nu_z e^{-y^2/w_z^2}+\frac 1 2  h\nu_x e^{-y^2/w_x^2} \label{eq:zpe}
\end{equation}
with the spin-dependent potential created by the near-resonant beam,
\begin{equation}
V_\sigma(y) = \epsilon_\sigma V_s e^{-2y^2/w_s^2} \label{eq:v_sigma}
\end{equation}
with $\epsilon_\uparrow = +1$ and $\epsilon_\downarrow = -1$.
This is the potential plotted in Figs.~\ref{fig:spin_gate}(d) to (f), neglecting the spatial variations of the attractive gate beam.
This real potential is furthermore used to determine the shift of the conductance curves in the Hartree mean-field model of Fig.~\ref{fig:interactions}, as explained in the following section.
Losses are included in the Landauer-B\"uttiker model displayed in Figs.~\ref{fig:spin_gate}(a) to (c) by adding an imaginary potential $iV_{\rm loss}(y) = -i\hbar \Gamma(y)/2$ proportional to the local scattering rate $\Gamma(y)$, see \cite{companion}.

\subsection*{Data analysis}

\subsubsection*{Conductance without temperature difference}
In the absence of a temperature difference $\Delta T = 0$, the particle current through the QPC is linear in the chemical potential difference between the reservoirs:
\begin{equation}
I_N = G \Delta \mu
\label{eq:particle_current}
\end{equation}
where $G$ is the conductance of the channel.
The linear approximation is valid for the weak interaction strengths considered here \cite{krinner_mapping_2016}. This further ensures that spin drag is negligible and that biases, currents, and transport coefficients can be treated independently for each spin $\sigma \in \{\uparrow, \downarrow\}$ (omitted in the rest of the subsection).

The chemical potential for each reservoir $r \in \{L, R\}$ and each spin can be furthermore expanded to first order around the atom number at $t = 0$, $d\mu_r = dN_r / \kappa_r$, where $\kappa_r$ is the compressibility of one reservoir.
Eq.~(\ref{eq:particle_current}) can then be simplified to a closed first order differential equation in the atom number difference $\Delta N$.
This difference is therefore an exponentially decreasing function of time with a time constant
\begin{equation}
\tau = G \left(\frac{1}{\kappa_{\rm L}}+\frac{1}{\kappa_{\rm R}} \right) \approx \frac {2G}{\kappa}
\label{eq:decay_time_RC}
\end{equation}
where we replace reservoir compressibilities at $t = 0$ by the reservoir compressibility $\kappa$ at global equilibrium for a mean atom number $\bar{N} = (N_L + N_R)/2$ and temperature $T$, without significant change in the results.
This represents the analogue of an RC circuit for neutral atoms, replacing charge by the number of atoms transferred from one reservoir to the other $\Delta N/2$, and capacitance by the compressibility per reservoir $\kappa$.

In Figs.~\ref{fig:spin_gate} and \ref{fig:interactions}, the time constant $\tau$ for each spin state is inferred from the atom number difference after a fixed transport time $t$
\begin{equation}
\tau = -t \, \ln \frac{\Delta N (t)}{\Delta N (0)}.
\label{eq:decay_time_Delta_N}
\end{equation}
We calibrate the initial value $\Delta N (0)$ for each set of experimental parameters (scattering length, near-resonant beam intensity and detuning) by averaging $\Delta N(t)$ for low values of the gate potential $V_g$, when transport is absent.
Together with the reservoir compressibility $\kappa$, Eq.~(\ref{eq:decay_time_RC}) allows us deduce the conductance $G$.

\subsubsection*{Extracting thermodynamic quantities from the cloud profile}

Extracting the conductance relies on the knowledge of the temperature, the atom number and the compressibility for each half-reservoir and each internal state.
To that end, we rely on the equation of state of non-interacting Fermi gases.
The atomic cloud is imaged after a short time-of-flight during which it only expands along the tight confinement axes $x, z$.
We integrate the density distribution along these axes and extract the second moment of the distribution along the weakly confined axis $y$.
This second moment can be related to the total energy of the gas thanks to the virial theorem for non-interacting Fermi gases.
The non-interacting equation of state allows to deduce the other relevant thermodynamic quantities such as chemical potential, temperature and reservoir compressibility.
We verified that the extracted quantities do not differ significantly from those obtained with an interacting equation of state in the mean-field limit \cite{su_low-temperature_2003} for the weak scattering lengths $|a| < 800\,a_0$ used here.

\subsubsection*{Fitting procedure}

The conductance shown in Fig.~\ref{fig:spin_gate} as a function of the local chemical potential $V_g + \mu_\text{res}$ is compared to a Landauer model presented and justified in \cite{companion} yielding the conductance $G_\sigma$ for each spin in the presence of dissipation.
The relevant model parameters are summarized by the following function
\begin{equation} 
y_\sigma(x) = A G_\sigma(x-\mu_0, T, \bar{\delta}, V_s), \label{eq:fit_function}
\end{equation}
where $A$ is a scaling factor capturing an effective decrease of the extracted conductance below $1/h$, $\mu_0$ is a chemical potential offset, $T$ is the temperature, $\bar{\delta}$ is the detuning relative to the mean resonance frequency and $V_s$ is the magnitude of the spin-dependent potential at $\bar{\delta} = 0$.

In Fig.~\ref{fig:spin_gate}, all points $(V_g + \mu_\text{res}, G_\sigma)$ associated with the same near-resonant beam intensity $I_s$ are fitted with (\ref{eq:fit_function}) for both spins $\sigma$ simultaneously.
We use $A$, $\mu_0$ and $V_s$ as free parameters, fix $T$ to the mean temperature independently measured from the cloud density profiles, and fix $\bar{\delta}$ to 0.
The fitted values of $V_s$ and fit errors are reported in the main text.
The fitted scale $A$ is about $0.8$ for all $I_s$ while the chemical potential offset $\mu_0$ is consistently below $k_B \cdot \SI{0.1}{\micro\kelvin}$, hinting at minor technical imperfections of the QPC potential or a small systematical error in the calibration of the QPC beams.

In Fig.~\ref{fig:interactions}(a), $(V_g + \mu_\text{res}, G_\sigma)$ is fitted for both spin states $\sigma$ simultaneously with the following function:
\begin{equation} 
y_\sigma(x) = \frac{A}{1 + \exp[-(x-\mu_{0, \sigma})/w]} + B \label{eq:fit_function_logistic}
\end{equation}
with a common amplitude $A$, width $w$, and vertical offset $B$, and independent chemical potential shifts $\mu_{0,\sigma}$ for the two spins.
The horizontal separation between the two curves is defined as $V_{s, \text{eff}} = \mu_{0, \uparrow} - \mu_{0, \downarrow}$.
In Fig.~\ref{fig:interactions}(b), $V_{s, \text{eff}}$ is then plotted versus the normalized beam intensity $I_s/I_\text{sat}$ and fitted with a linear function $y(x) = C x$.

All fits described above are performed by orthogonal distance regression \cite{boggs_orthogonal_1990} to include errors along $x$ and $y$.

\section*{Hartree mean-field model}

Here, we detail how we model interaction effects on transport through the QPC at the mean-field level. In this description, a single particle is effectively embedded in the mean density of the other particles. Interactions introduce a shift of the single-particle energies which is proportional to the density and is used to calculate conductances within the Landauer framework.

\subsection*{Mean-field Hamiltonian}

First we derive a mean-field Hamiltonian and self-consistent equations to calculate the densities inside the near-resonant tweezer. As we focus on the center of the QPC, the transverse confinements and spin-dependent potential are uniform along the direction of transport. This leads to the following Hamiltonian consisting of kinetic and potential energy in second quantization 
\begin{equation}
    H_0 = \sum_\sigma \int \psi_\sigma^\dag (\mathbf{r}) 
    \left( - \frac{\hbar^2 \Delta}{2 m} + V_\sigma (\mathbf{r}) \right) 
    \psi_\sigma (\mathbf{r}) \, \mathrm{d^3} \mathbf{r}, 
\end{equation}
where the field operator $\psi_\sigma (\mathbf{r})$ annihilates a particle with spin~$\sigma$ at position~$\mathbf{r}$. The potential is given by 
\begin{equation}
    V_\sigma (\mathbf{r}) = \frac{1}{2} m \omega_x^2 x^2 + \frac{1}{2} m \omega_z^2 z^2 + V_\sigma
\end{equation}
where $\omega_{x,z} = 2 \pi \cdot \nu_{x,z}$.
The interparticle interactions for ultracold atoms are of the contact type and captured by the Hamiltonian
\begin{equation}
    H_\text{int} = \frac{g}{2} \sum_{\sigma} \int 
    \psi_{\sigma}^\dag (\mathbf{r}) \psi_{\bar{\sigma}}^\dag (\mathbf{r}) 
    \psi_{\bar{\sigma}} (\mathbf{r}) \psi_{\sigma} (\mathbf{r}) \, \mathrm{d^3} \mathbf{r}
\end{equation}
with the interaction constant $g = 4 \pi \hbar^2 a / m$ at scattering length~$a$. As both spin species are summed the prefactor is necessary to avoid double counting. Due to Pauli's principle a particle with spin $\sigma$ only interacts with one of opposite spin $\bar{\sigma}$.

\subsubsection*{Basis change}
As the channel provides harmonic confinement in transverse directions and free motion in longitudinal direction, it is convenient to change the basis states via 
\begin{equation}
    \psi_\sigma (\mathbf{r}) = \sum_{\mathbf{n} k} 
    \phi_{n_x} (x) \phi_{n_z} (z) \frac{1}{\sqrt{L}} e^{i k y} a_{\mathbf{n} k \sigma}.
\end{equation}
Here, we label the harmonic transverse modes with a multi-index $\mathbf{n} = (n_x, n_z)$ and the longitudinal plane wave with the wavenumber $k$ that is a multiple of $2 \pi / L$. The harmonic eigenstates along the $x$ axis are 
\begin{equation}
    \phi_{n_x} (x) = \frac{1}{\left( \pi a_x^2 \right)^{1/4}} \frac{1}{\sqrt{2^{n_x} n_x !}} \cdot 
    e^{- x^2 / 2 a_x^2} \cdot H_{n_x} (x / a_x)
\end{equation}
with the Hermite polynomials $H_n$ of order $n$ and the natural length $a_x = \sqrt{\hbar / m \omega_x}$. The formula reads analogously along the $z$ axis.

Now, we rewrite the Hamiltonians $H_0$ and $H_\text{int}$ in this basis and only keep the lowest transverse mode $\mathbf{n} = \mathbf{0}$ as we work in the single-mode regime. Then, the Hamiltonian $H_0$ operating on the single-particle space becomes 
\begin{equation}
    H_0^\text{sm} = \sum_{k \sigma} 
    \left[ \frac{\hbar^2 k^2}{2 m} + V_0 + V_{\sigma} \right] 
    a_{\mathbf{0} k \sigma}^\dag a_{\mathbf{0} k \sigma}
\end{equation}
with the energy $V_0 = \hbar \omega_x / 2 + \hbar \omega_z / 2$ in the transverse direction. The interaction Hamiltonian $H_\text{int}$ is cast into 
\begin{multline}
    H_\text{int}^\text{sm} = \frac{g}{2} 
    \sum_{\sigma} 
    \sum_{\substack{k_1 k_2 \\ k_3 k_4}} 
    \frac{\text{ov}}{a_x a_z} \frac{1}{L} \delta_{k_1 + k_2, k_3 + k_4} \times \\
    \times \, a_{\mathbf{0} k_1 \sigma}^\dag a_{\mathbf{0} k_2 \bar{\sigma}}^\dag 
    a_{\mathbf{0} k_3 \bar{\sigma}} a_{\mathbf{0} k_4 \sigma}.
\end{multline}
The spatial integration along $y$ reduces to the Kronecker delta $\delta_{k_1 + k_2, k_3 + k_4}$ which ensures momentum conservation. The integration over the $x$ and $z$ directions gives rise to the overlap
\begin{equation}
    \text{ov} = 
    \int_{-\infty}^{\infty} \varphi_{0}^4 (\xi) \, \mathrm{d} \xi \cdot 
    \int_{-\infty}^{\infty} \varphi_{0}^4 (\xi) \, \mathrm{d} \xi = \frac{1}{2 \pi}
\end{equation}
with the dimensionless wavefunction $\varphi_0 (\xi) = \sqrt{a_x} \cdot \phi_0 (\xi a_x)$ and reduced coordinates $\xi = x / a_x$, and analogously along the $z$ axis. 

As we focus on a single transverse mode and elastic collisions are restricted to 1D the individual particle momenta are preserved along $y$ ($k_1 = k_4$, $k_2 = k_3$). Hence, the interaction Hamiltonian simplifies to 
\begin{equation}
    H_\text{int}^\text{sm} = \frac{g}{2} 
    \sum_{\sigma} 
    \sum_{k_1 k_2} 
    \frac{\text{ov}}{a_x a_z} \frac{1}{L} \,
    a_{\mathbf{0} k_1 \sigma}^\dag a_{\mathbf{0} k_2 \bar{\sigma}}^\dag 
    a_{\mathbf{0} k_2 \bar{\sigma}} a_{\mathbf{0} k_1 \sigma}.
\end{equation}
As the operators $a_\alpha$ and $a_\alpha^\dag$ obey anti-commutation relations the product of operators in the Hamiltonian can be brought into the form $a_\alpha^\dag a_\alpha a_\beta^\dag a_\beta$. By applying the mean-field approximation in the density operators $a_\alpha^\dag a_\alpha$ and $a_\beta^\dag a_\beta$ we obtain 

\begin{equation}
    H_\text{int}^\text{MF} = 
    \sum_{\sigma} \sum_{k} 
    \left( \frac{\hbar^2 k^2}{2 m} + \varepsilon_{\sigma} \right) 
    a_{\mathbf{0} k \sigma}^\dag a_{\mathbf{0} k \sigma} + E_0
    \label{eq:mean_field_hamiltonian}
\end{equation}
with the energy $\varepsilon_{\sigma}$ that includes the transverse energy, the spin-dependent potential, and the interaction shift 
\begin{equation}
    \varepsilon_{\sigma} = 
    V_0 + V_{\sigma} + 
    \frac{1}{L} \sum_{k^\prime} U
    \langle a_{\mathbf{0} k^\prime \bar{\sigma}}^\dag a_{\mathbf{0} k^\prime \bar{\sigma}} \rangle
    \label{eq:single_particle_energy_v1}
\end{equation}
and an energy offset $E_0$ that avoids overcounting: 
\begin{equation}
    E_0 = - \frac{1}{2 L} \sum_{\sigma} 
    \sum_{k k^\prime} U 
    \langle a_{\mathbf{0} k \sigma}^\dag a_{\mathbf{0} k \sigma} \rangle
    \langle a_{\mathbf{0} k^\prime \bar{\sigma}}^\dag a_{\mathbf{0} k^\prime \bar{\sigma}} \rangle.
\end{equation}
Here, the coupling constant is $U = g \cdot \text{ov} / (a_x a_z)$. The mean-field Hamiltonian is diagonal in the occupation operator $a_\alpha^\dag a_\alpha$ which simplifies the following treatment.

\subsection*{Density calculation}

Based on the mean-field Hamiltonian~\eqref{eq:mean_field_hamiltonian} we derive formulae for the mean density and its variance and pinpoint when particles start to enter the tweezer. With these expressions we discuss the situation at our experimental parameters.

\subsubsection*{Mean density}

The line density of each spin state is obtained from the occupation numbers by summing over the wavenumber.
\begin{equation}
    n_{\sigma} = \frac{1}{L} \sum_k \langle a_{\mathbf{0} k \sigma}^\dag a_{\mathbf{0} k \sigma} \rangle
    \label{eq:line_density}
\end{equation}
For a diagonal Hamiltonian the occupation number is given in the grand canonical ensemble by 
\begin{equation}
    \langle a_{\mathbf{0} k \sigma}^\dag a_{\mathbf{0} k \sigma} \rangle = 
    f \left( \beta \left[ \frac{\hbar^2 k^2}{2 m} + \varepsilon_{\sigma} - \mu_\sigma \right] \right)
\end{equation}
with the Fermi-Dirac distribution $f(x) = 1 / (1 + e^x)$, the inverse temperature $\beta = 1 / k_B T$ and the chemical potential $\mu_\sigma$ of spin state $\sigma$ imposed by the reservoirs. In the thermodynamic limit ($L \rightarrow \infty$), the summation becomes an integral over the wavenumber which can be transformed into an integral over energy to arrive at the expression 
\begin{equation}
    n_{\sigma} = \frac{1}{\lambda} F_{-1/2} ( \beta \left[ \mu_\sigma - \varepsilon_{\sigma} \right] ), 
    \label{eq:equation_of_state}
\end{equation}
where $F_j$ denotes the complete Fermi-Dirac integral of order $j$ \cite[Eq. (25.12.14)]{Olver2010} and $\lambda$ the thermal wavelength given by 
\begin{equation}
    \lambda = \sqrt{\frac{2 \pi \hbar^2}{m k_B T}}.
\end{equation}
The energy $\varepsilon_{\sigma}$ defined in \eqref{eq:single_particle_energy_v1} can be written with equation~\eqref{eq:line_density} as 
\begin{equation}
    \varepsilon_{\sigma} = 
    V_0 + V_{\sigma} + 
    U n_{\bar{\sigma}}.
\end{equation}
The single-particle energy $\varepsilon_{\sigma}$ depends on the density of the opposite spin and, together with the equation of state \eqref{eq:equation_of_state}, forms a system of equations for the line density that needs to be solved self-consistently.

\subsubsection*{Density variance}

To estimate the fluctuations around the mean line density we start with the particle number fluctuations in the grand canonical ensemble \cite{Stephenson1974} that is valid for all system sizes: 
\begin{equation}
    \langle N_\sigma^2 \rangle - \langle N_\sigma \rangle^2 = 
    k_B T \diff*{\langle N_\sigma \rangle}{\mu_\sigma}{T}.
\end{equation}
The particle number fluctuations are converted into fluctuations of the densities via $\hat{N}_\sigma = L \hat{n}_\sigma$ giving 
\begin{equation}
    \langle n_\sigma^2 \rangle - \langle n_\sigma \rangle^2 = 
    \frac{k_B T}{L} \diff*{\langle n_\sigma \rangle}{\mu_\sigma}{T}.
    \label{eq:density_fluctuations}
\end{equation}
It is visible that fluctuations are stronger with increasing temperature and for smaller systems. The variation of densities with chemical potential contains the fermionic particle statistics and the interparticle interactions.

\subsubsection*{Density onset at zero temperature}
At zero temperature, equation \eqref{eq:equation_of_state} simplifies to
\begin{equation}
    n_{\sigma} = \sqrt{\frac{2 m}{\pi^2 \hbar^2}} \cdot 
    \theta(\mu - \varepsilon_{\sigma}) \cdot 
    \sqrt{\mu - \varepsilon_{\sigma}}
    \label{eq:density_zero_temp}
\end{equation}
with the Heaviside step function $\theta$ and chemical potential $\mu$ equal for both spin states ($\mu_\downarrow = \mu_\uparrow$). The density displays the square root behavior in chemical potential expected from the $1/\sqrt{E}$ proportionality of the one-dimensional system's density of states. At small chemical potentials, only the state $\ket{\downarrow}$ attracted by the tweezer occupies the channel and hence its density is unaffected by interactions. It reads 
\begin{equation}
    n_{\downarrow} = \sqrt{\frac{2 m}{\pi^2 \hbar^2}} \cdot 
    \theta(\mu - V_0 - V_\downarrow) \cdot 
    \sqrt{\mu - V_0 - V_\downarrow}.
    \label{eq:density_down_zero_temp}
\end{equation}
Clearly, the particles in state $\ket{\downarrow}$ start to occupy the channel at the chemical potential $\acute{\mu}_\downarrow = V_0 + V_\downarrow$. The threshold of the other state depends on interactions and follows from formula \eqref{eq:density_zero_temp} via $\acute{\mu}_\uparrow - \varepsilon_{\uparrow} = 0$ that is  
\begin{equation}
    \acute{\mu}_\uparrow - V_0 - V_\uparrow - U n_{\downarrow} = 0.
\end{equation}
Together with equation \eqref{eq:density_down_zero_temp} this forms a self-consistent equation for $\acute{\mu}_\uparrow$. The solution is 
\begin{align}
    \acute{\mu}_\uparrow &= V_0 + V_\uparrow + \Delta \mu_\text{int}, \\
    \Delta \mu_\text{int} &= 2 u^2 + 2 u \sqrt{u^2 + V_\uparrow - V_\downarrow}
    \label{eq:interaction_shift}
\end{align}
with the scaled coupling constant 
\begin{equation}
    u = \sqrt{\frac{m}{2 \pi^2 \hbar^2}} U.
\end{equation}

\begin{figure*}
    \includegraphics{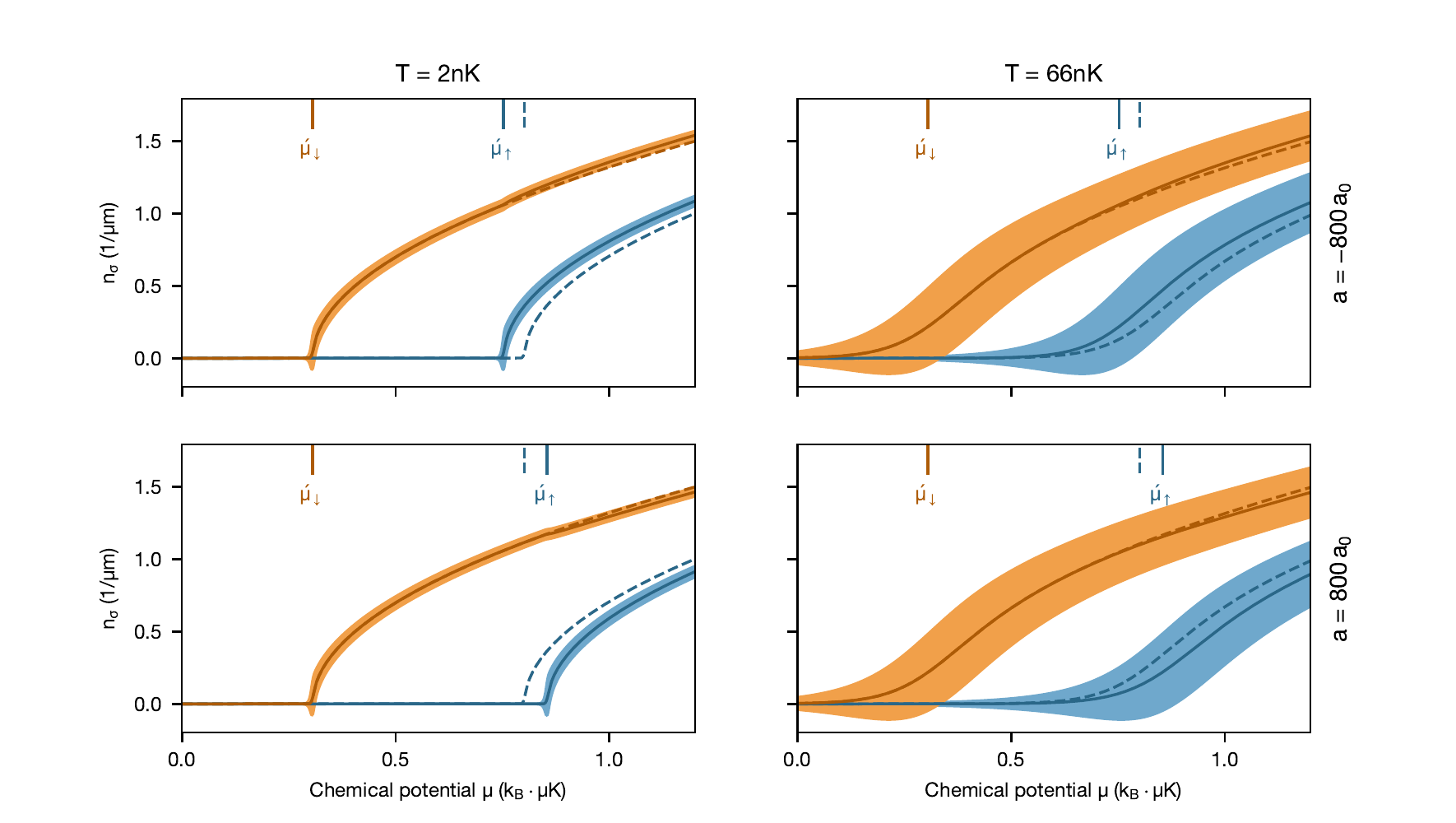}
    \caption{\textbf{Self-consistent line density.} Density $n_\sigma$ as a function of chemical potential $\mu$ equal for both spin states at different temperatures $T$ and scattering lengths $a$ ($\ket{\downarrow}$ in orange and $\ket{\uparrow}$ in blue). The shaded regions indicate the standard deviation \eqref{eq:density_fluctuations} estimated in the grand canonical ensemble for a system size $L = w_s = \SI{2.0}{\micro\metre}$. For comparison the non-interacting predictions are shown at the same temperature (dashed curves). The vertical lines indicate the chemical potential onsets $\acute{\mu}_\sigma$ at zero temperature with (solid line) and without interactions (dashed line). The spin-dependent potential $V_\sigma$ is $k_B \cdot \SI{\pm 248}{\nano\kelvin}$ and the transverse trapping frequencies are $\omega_x = 2\pi \cdot \SI{14.0}{\kilo\hertz}$ and $\omega_z = 2\pi \cdot \SI{9.03}{\kilo\hertz}$.}
    \label{fig:density_vs_mu}
\end{figure*}

\subsubsection*{Results and discussion}

Figure~\ref{fig:density_vs_mu} displays the density $n_\sigma$ versus chemical potential~$\mu$ for both spin states at two temperatures and scattering lengths obtained using equation \eqref{eq:equation_of_state}. The non-interacting predictions are presented as dashed curves and, as expected, they follow a square root behavior at low temperatures. At the chemical potential $\acute{\mu}_\downarrow$, particles attracted by the tweezer start to enter the channel. Its location is independent of interactions as particles of the other state are absent. This is valid if the effective Zeeman splitting avoids thermal occupation, meaning when $V_\uparrow - V_\downarrow \gg k_B T$. The onset $\acute{\mu}_\uparrow$ where the other state occupies the tweezer is shifted towards smaller (larger) chemical potentials with attractive (repulsive) interactions. The shift is quantified by $\Delta \mu_\text{int}$ in equation \eqref{eq:interaction_shift} that is asymmetric in interactions. It is larger in the repulsive regime ($a > 0$) than in the attractive case with opposite scattering length $-a$. That is because the energies $\varepsilon_{\sigma}$ change linearly with the densities while the densities behave nonlinearly with chemical potential. The asymmetry is typically small for our experimental parameters. For example in Fig.~\ref{fig:density_vs_mu} the shifts $\Delta \mu_\text{int}$ are $k_B \cdot \SI{54}{\nano\kelvin}$ for a scattering length of $800 \, a_0$ and $-k_B \cdot \SI{48}{\nano\kelvin}$ for $-800 \, a_0$ respectively. Besides the onset locations, the densities themselves also deviate from the non-interacting expectation with interactions when both spin states are present in the tweezer. Attractive interactions increase the densities while they are reduced with repulsion.

The density fluctuations are indicated with the standard deviation (shaded regions) and are larger at higher temperatures. They are useful to check if they are indeed small as assumed by the mean-field theory. At our typical temperature of \SI{66}{\nano\kelvin} the fluctuations only dominate at small chemical potentials and are moderate compared to the mean density when particles enter the tweezer. Hence despite the small system size we expect mean-field theory to be valid within the interesting chemical potential range.

\begin{figure}
    \includegraphics{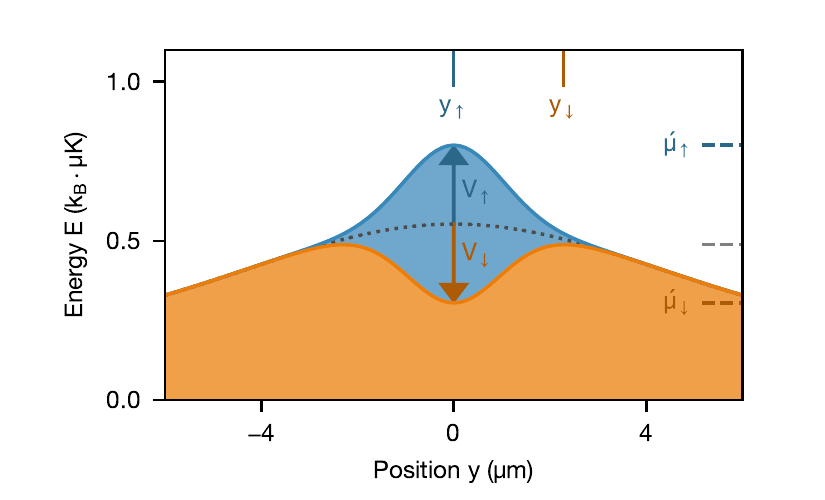}
    \caption{\textbf{Effective one-dimensional potential.} The horizontal colorful lines indicate the density onsets $\acute{\mu}_\sigma$ at zero temperature and without interactions and in gray the shifted conductance transition due to the effective potential. The experimental parameters are as for Fig.~\ref{fig:density_vs_mu}.}
    \label{fig:eff_pot}
\end{figure}

\subsection*{Conductance calculation}

In the following we determine conductances including interactions via the modified densities. We start with the Landauer formula assuming infinitesimal bias between the reservoirs and classical transmission through the QPC: 
\begin{equation}
    G_\sigma = \frac{1}{h} f ( \beta \left[ \mu_\sigma - \wideparen{\varepsilon}_{\sigma} \right] ),
\end{equation}
where $f(x) = (1+e^{-x})^{-1}$ is the inverted Fermi-Dirac distribution.
The conductances are related to the single-particle energies $\wideparen{\varepsilon}_{\sigma}$ that are modified by interactions as $\varepsilon_{\sigma}$ but additionally consider the non-locality of transport. While the energies $\varepsilon_{\sigma}$ are related to the center, transport is dominated by the maxima in the effective potential that may be off-centered, as shown in Fig.~\ref{fig:eff_pot}. Assuming everywhere the same density as at the center, the doubly-modified single-particle energies are 
\begin{equation}
    \wideparen{\varepsilon}_{\sigma} = 
    \wideparen{V}_0 + \wideparen{V}_{\sigma} + 
    U n_{\bar{\sigma}} \\
\end{equation}
where the transverse energy $\wideparen{V}_0$ and spin-dependent potential $\wideparen{V}_{\sigma}$ are obtained by evaluating (\ref{eq:zpe}) and (\ref{eq:v_sigma}) at the maxima locations $y_{\sigma}$.

\begin{figure*}
    \includegraphics{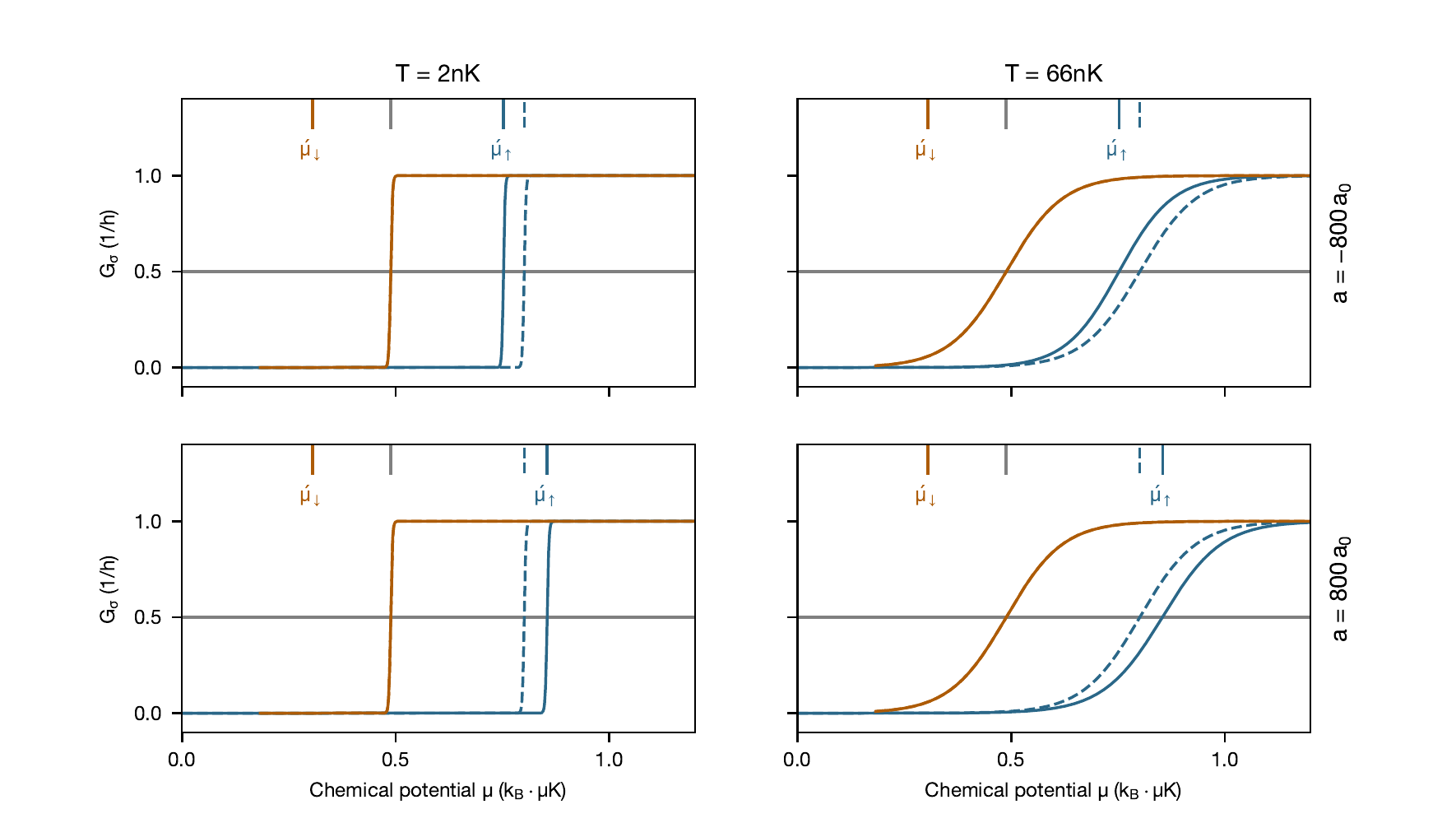}
    \caption{\textbf{Conductance with mean-field interactions.} Conductance $G_\sigma$ versus chemical potential $\mu$ equal for both spin states at different temperatures $T$ and scattering lengths $a$ ($\ket{\downarrow}$ in orange and $\ket{\uparrow}$ in blue). Non-interacting predictions are shown at the same temperature (dashed curves). The vertical colorful lines indicate the density onsets $\acute{\mu}_\sigma$ at zero temperature with (solid line) and without interactions (dashed line) and in gray the shifted onset due to the effective potential. The experimental parameters are as for Fig.~\ref{fig:density_vs_mu}.}
    \label{fig:mft_conductance}
\end{figure*}

\subsubsection*{Results and discussion}

Fig.~\ref{fig:mft_conductance} shows the conductance $G_\sigma$ versus chemical potential $\mu$ for the same conditions as for the densities in graph~\ref{fig:density_vs_mu}. Upon increasing the chemical potential the attracted state $\ket{\downarrow}$ starts to occupy the tweezer beyond the density onset $\acute{\mu}_\downarrow$ (vertical blue line). However, due to the shape of the effective potential, shown in Fig.~\ref{fig:eff_pot}, conductance rises later and the transition is centered around the vertical gray line. As expected, the transition of state $\ket{\uparrow}$ agrees with the density onset $\acute{\mu}_\uparrow$ and is modified by interactions. For comparison the non-interacting predictions are shown at the same temperatures (dashed curves). From the conductance curves the chemical potential separation between the two states is extracted at half conductance quantum $1/2h$, as plotted in Fig.~\ref{fig:interactions}. 

Additionally to interactions, dissipation influences the separation as well. In principle it can be included in the mean-field description using for example the quantum jump approach \cite{Molmer1993}. Here, we discuss dissipation qualitatively considering only its effect on the densities. The particle losses will overall reduce conductances but the location of the first transition will remain fixed as the chemical potential is imposed by the reservoirs and interactions have no effect. At the second transition the density $n_\downarrow$ is reduced by dissipation, which should also reduce the absolute shift compared to the non-interacting case.

\bibliographystyle{apsrev4-1}
% \bibliography{paper}

\end{document}